\documentclass{appolb}

\usepackage{amsmath}
\usepackage{amssymb}
\usepackage{cite}
\usepackage[left=2.5cm,right=2.5cm,top=2.cm,bottom=2.3cm]{geometry}
\usepackage{graphicx}
\usepackage[font=footnotesize]{caption}
\usepackage[font=footnotesize]{subcaption}
\usepackage{xcolor}
\usepackage[colorlinks=true,linkcolor=blue]{hyperref}
\usepackage[capitalise]{cleveref}

\newcommand{\C}{\mathbb{C}}
\newcommand{\ii}{\mathrm{i}}
\newcommand{\SL}{\mathrm{SL}}
\newcommand{\SU}{\mathrm{SU}}

\DeclareMathOperator{\tr}{tr}

\graphicspath{{./figures}}

\begin{document}

	\title{Equation of state of hot and dense QCD using the complex Langevin equation%
	\thanks{Presented at Confinement 2026.}}
	\author{Michael Mandl, Dénes Sexty, Daniel Unterhuber
	\address{Institute of Physics, NAWI Graz, University of Graz, Universitätsplatz 5, 8010 Graz, Austria}}
	\maketitle
	
	\begin{abstract}
		We present continuum-extrapolated lattice results for the QCD equation of state at high temperatures and unprecedentedly large baryon chemical potentials. These results were obtained by performing the first complex Langevin simulations at the physical value of the pion mass. We work in a strangeness-neutral, charm-neutral, and isospin-symmetric setup and show results for the baryon density, pressure, and trace anomaly as a function of baryon chemical potential and temperature, respectively.
	\end{abstract}
	
	\section{Introduction}
	Quantum chromodynamics (QCD) under extreme conditions, such as high temperature and baryon number density, is the theoretical framework underlying various physical systems of contemporary interest, such as neutron stars, heavy-ion collisions, and the early universe. A particularly prominent question concerns the nature of the phase diagram of QCD in the plane spanned by the baryon chemical potential $\mu_B$ and the temperature $T$. While the $T$-axis of this phase diagram is understood quite well from a theoretical point of view, much less is known about the non-zero-density region. A major reason for this is that lattice quantum field theory, to date the most successful tool for the non-perturbative description of strongly-interacting theories from first principles, cannot be applied straightforwardly at non-zero baryon chemical potential due to the infamous sign problem preventing conventional lattice techniques based on importance sampling \cite{AS26r}.

One approach that is unaffected by the sign problem and allows for direct simulations of finite-density QCD in principle is the complex Langevin method \cite{Par83,Kla83}, an extension of Parisi--Wu stochastic quantization \cite{PW81} to theories with complex actions. The method is based on the simulation of a stochastic process on a complexified field manifold and has shown promising success in the high-temperature region of QCD at non-zero baryon density \cite{Sex14,Sex14_2,ASS14,FKS15,ASS17,NNS18,Sex19,KS19,IMN20,SSS20_2,
AJZ20r,AJZ22u,HS24,TAI25}, most recently resulting in the first ever complex Langevin study at the physical point \cite{MSU26}, computing the baryon density and pressure of QCD for ratios of baryon chemical potential to temperature up to $\mu_B/T\sim10$, albeit at rather high temperatures in the range $260\,\mathrm{MeV}\leq T\leq600\,\mathrm{MeV}$. In this contribution, we provide a complimentary view on these results and furthermore discuss the trace anomaly as well.
	
	\section{Complex Langevin dynamics}
	In the context of lattice QCD, the complex Langevin approach entails the complexification of the usual $\SU(3)$-valued gauge links to non-compact degrees of freedom $U_{x\mu}\in\SL(3,\C)$. The gauge-link configurations (to be averaged over in a standard statistical analysis) are then generated according to an evolution in an auxiliary time direction $\tau$ called the Langevin time. More precisely, the $\tau$-evolution is governed by a generalization of the complex Langevin equation \cite{Par83,Kla83} to group-valued variables. In its discretized form, a straightforward update step propagating the gauge links $U_{x\mu}^{(\tau)}$ at Langevin time $\tau$ forward to Langevin time $\tau+\Delta\tau$, with $\Delta\tau>0$ denoting a discrete step-size, can be written as follows:
\begin{equation}\label{eq:cle}
	U_{x\mu}^{(\tau+\Delta\tau)} = \exp\left[\ii\sum_{a=1}^{8}\lambda_a\left(-\Delta\tau D_{ax\mu}S[U] + \sqrt{2\Delta\tau}\,\eta_{ax\mu}^{(\tau)}\right)\right]U_{x\mu}^{(\tau)}\;.
\end{equation}
Here, $\lambda_a$ denotes the usual Gell-Mann matrices, $S[U]=S_g[U]-\ln\det M[U]$ is the (Euclidean) effective action of QCD after integrating out the fermions, with the gauge action $S_g$ and fermion matrix $M$, $D_{ax\mu}$ denotes the left derivative, defined by its action on functions $f[U]$ of the gauge links,
\begin{equation}
	D_{ax\mu}f[U] = \frac{\partial}{\partial\alpha} f\left[e^{\ii\alpha\lambda_a}U_{x\mu}\right]\bigg\vert_{\alpha=0}\;,
\end{equation}
and $\eta_{ax\mu}^{(\tau)}$ are real Gaussian noise variables obeying 
\begin{equation}
	\left\langle\eta_{ax\mu}^{(\tau)}\right\rangle=0 
	\quad \textnormal{and} \quad
	\left\langle\eta_{ax\mu}^{(\tau)}\eta_{by\nu}^{(\tau')}\right\rangle=\delta_{ab}\delta_{xy}\delta_{\mu\nu}\delta_{\tau\tau'}\;.
\end{equation}
In the present study, however, we use an improved version of \eqref{eq:cle} \cite{UF85}, as outlined in \cite{MSU26}, in order to reduce finite-step-size effects. Moreover, we employ an adaptive-step-size algorithm \cite{AJS10} and the gauge-cooling procedure \cite{SSS13,NNS16_2} to stabilize the complex Langevin evolution. In our simulations, we have not encountered excursions of the evolution trajectory far into the non-unitary domain (as measured by the unitarity norm \cite{SSS20_2}), such that we are confident that our results are not affected by the wrong-convergence problem (see, e.g., \cite{Man26}).
	
	\section{Lattice setup}
	In our large-scale lattice simulations of full QCD at non-zero temperature and baryon density using the complex Langevin method, we study $2+1+1$ flavors of staggered quarks and employ a tree-level improved Symanzik gauge action. Moreover, we perform four steps of stout smearing \cite{MP04,Sex19}. We consider lattices of size $N_s^3\times N_t$, with $N_t$ and $N_s$ denoting the number of lattice points in the temporal and each spatial direction, respectively. Our lattices have an aspect ratio of $N_s/N_t=2$ and we consider three different temporal extents, $N_t\in\{10,12,16\}$. Our setup follows closely that of \cite{BBF15}, from which we also take the line of constant physics (LCP) along which we extrapolate our results to the continuum limit. While our spatial lattice sizes are smaller than those considered in \cite{BBF15}, we find finite-volume effects to be under control via explicit cross-checks with larger volumes, yielding consistent results within the statistical uncertainties.

In order to simulate QCD at strangeness and charm neutrality and to ensure an isospin-symmetric setting, we employ chemical potentials of the form
\begin{equation}
	\mu_u=\mu_d=:\mu_l\;, \quad \mu_s=0\;, \quad \mu_c=0\;,
\end{equation}
for the up ($u$), down ($d$), strange ($s$), and charm ($c$) quark, respectively. For the chemical potentials corresponding to the conserved charges of baryon number ($B$), electric charge ($Q$), strangeness ($S$), and charm ($C$), this choice entails
\begin{equation}
	\mu_B = 3\mu_l\;, \quad \mu_S = \mu_l = -\mu_C \;, \quad \mu_Q = 0\;, 
\end{equation}
and leads to vanishing strangeness and charm densities in our simulations to a very good approximation, as we have explicitly verified. For more details on our lattice setup, we refer to \cite{MSU26}. 
	
	\section{QCD equation of state}
	In this section, we present our results for the QCD equation of state at non-zero chemical potential in the high-temperature regime. Throughout, quantities that are normalized by appropriate powers of the temperature to make them dimensionless are indicated by hats. In particular, we define
\begin{equation}
	\hat{\mu}_B = \frac{\mu_B}{T}\;, \quad
	\hat{n}_B = \frac{n_B}{T^3}\;, \quad
	\hat{p} = \frac{p}{T^4}\;, \quad
	\hat{I} = \frac{I}{T^4}\;,
\end{equation}
for the baryon chemical potential, baryon number density, pressure, and trace anomaly, respectively.

In \cite{MSU26}, we plotted $\hat{n}_B$ and $\hat{p}$ as a function of $\hat{\mu}_B$ for different temperatures above the QCD crossover and compared with results from hard-thermal-loop perturbation theory (HTLpt) \cite{HBA14}, as well as with lattice simulations using a more conventional hybrid-Monte-Carlo (HMC) setup and a resummed Taylor expansion to compute the equation of state at non-zero chemical potential \cite{BGK22}. Indeed, we found good overall agreement with both HTLpt and HMC in the small-$\hat{\mu}_B$ region where all three methods are believed to be reliable. In what follows, we show $\hat{n}_B$ and $\hat{p}$ also as a function of the temperature for different $\hat{\mu}_B$ and we present results for the trace anomaly as well.

\begin{figure}[t]
	\centerline{
		\begin{subfigure}{.49\linewidth}
			\includegraphics[scale=0.48]{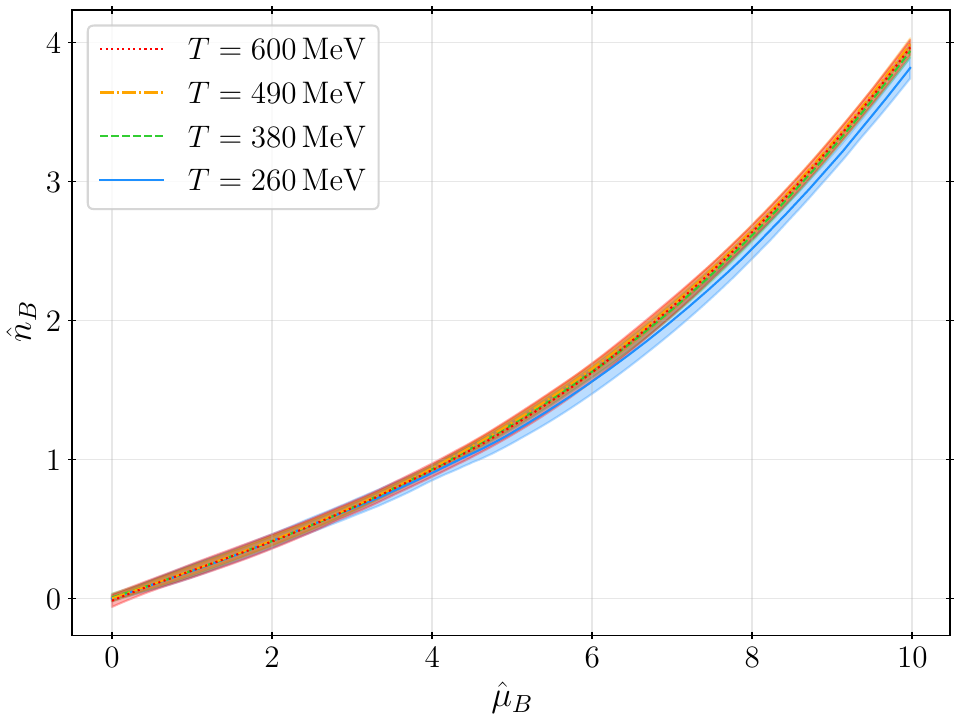}
		\end{subfigure}
		\begin{subfigure}{.49\linewidth}
			\includegraphics[scale=0.48]{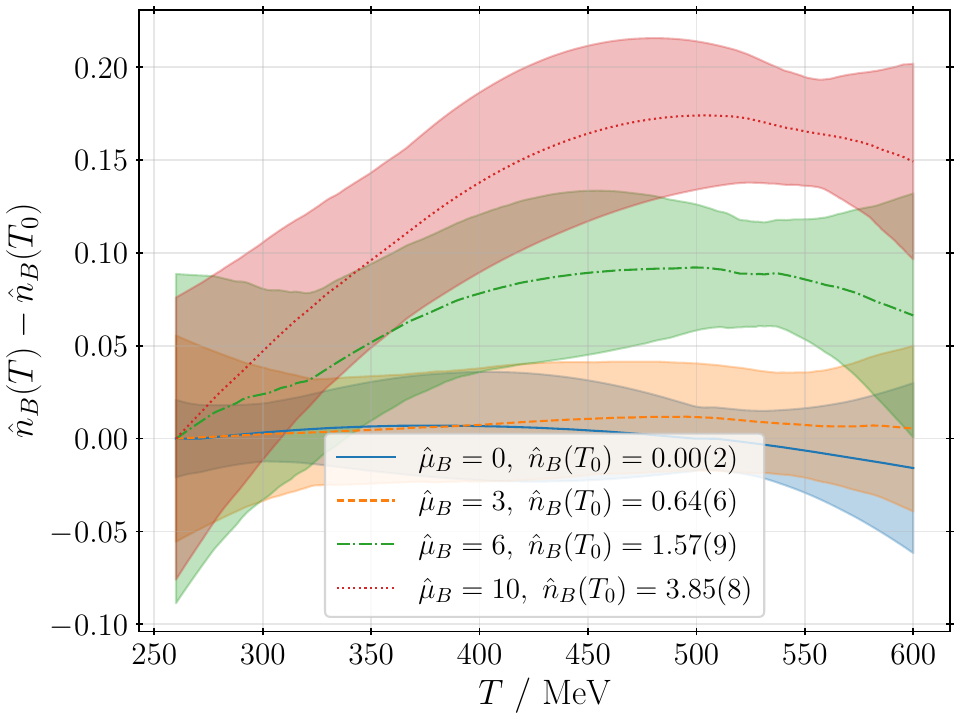}
		\end{subfigure}
	}
	\caption{Baryon number density on different lines within the $(\mu_B,T)$-plane. \emph{(left)}: $\hat{n}_B$ as a function of $\hat{\mu}_B$ for different temperatures. \emph{(right)}: $\hat{n}_B(T)-\hat{n}_B(T_0)$, with $T_0=260\,\mathrm{MeV}$, as a function of $T$ for different $\hat{\mu}_B$.}
	\label{fig:n_B}
\end{figure}

For the present analysis, the only quantity we actually need to measure on the lattice is the baryon number density, defined as
\begin{equation}
	n_B = \frac{T}{V}\frac{\partial\ln\mathcal{Z}}{\partial\mu_B} = \frac{T}{12V}\sum_{f=u,d,s,c}\left\langle\tr\left(M_f^{-1}\frac{\partial M_f}{\partial\mu_f}\right)\right\rangle\;,
\end{equation}
where $M_f$ is the Dirac matrix for the fermion flavor $f$ and $V$ denotes the lattice volume in physical units. The $\hat{\mu}_B$-dependence of $\hat{n}_B$ for different temperatures is shown in \cref{fig:n_B} (left). Note that for this result a fit of $\hat{n}_B$ in $\hat{\mu}_B$ was performed before the continuum extrapolation (see \cite{MSU26} for details on the fit and error analysis). One observes the expected increase of $n_B$ with the chemical potential and only a mild dependence of $\hat{n}_B$ on the temperature, especially for small $\hat{\mu}_B$, suggesting a predominantly cubic behavior, $n_B(T)\propto T^3$, in this region if $\hat{\mu}_B$ is kept fixed. It should be noted that the largest values of $\hat{\mu}_B$ we can reliably reach in our simulations, $\hat{\mu}_B\lesssim10$, are significantly higher than what is possible in HMC-based approaches, where the maximum ratio of baryon chemical potential to temperature is typically at most $3-4$. In \cite{MSU26}, we have discussed the role of saturation effects in our continuum extrapolation in some detail, whereas in the present work we restrict to the region in $\hat{\mu}_B$ in which saturation effects are negligible.

In \cref{fig:n_B} (right), we show the subtracted dimensionless density $\hat{n}_B(T)-\hat{n}_B(T_0)$ as a function of $T$ for different $\hat{\mu}_B$. Here, $T_0=260\,\mathrm{MeV}$ is the lowest temperature we have access to in our simulations. The reason for studying the subtracted quantity is to better resolve the $T$-dependence of $\hat{n}_B$, which is indeed approximately constant within the combined statistical and systematic uncertainties for small values of $\hat{\mu}_B$, confirming the above observation, while deviations become visible for larger $\hat{\mu}_B$. For completeness, the respective values of $\hat{n}_B(T_0)$ (whose uncertainties are ignored in the subtraction) are also indicated in \cref{fig:n_B} (right). 

\begin{figure}[t]
	\centerline{
		\begin{subfigure}{.49\linewidth}
			\includegraphics[scale=0.48]{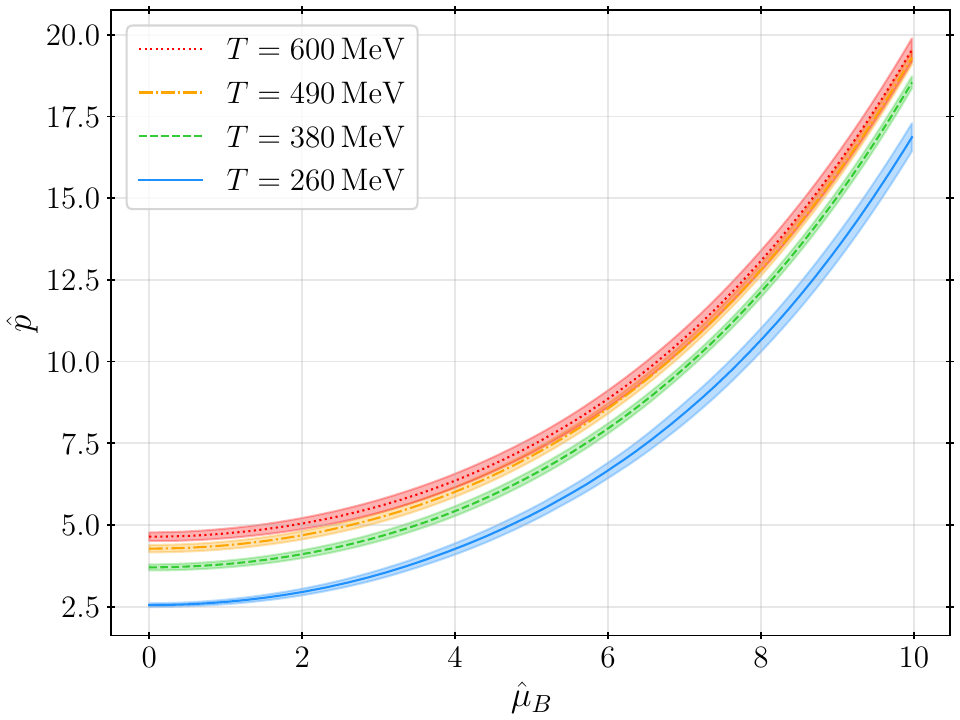}
		\end{subfigure}
		\begin{subfigure}{.49\linewidth}
			\includegraphics[scale=0.48]{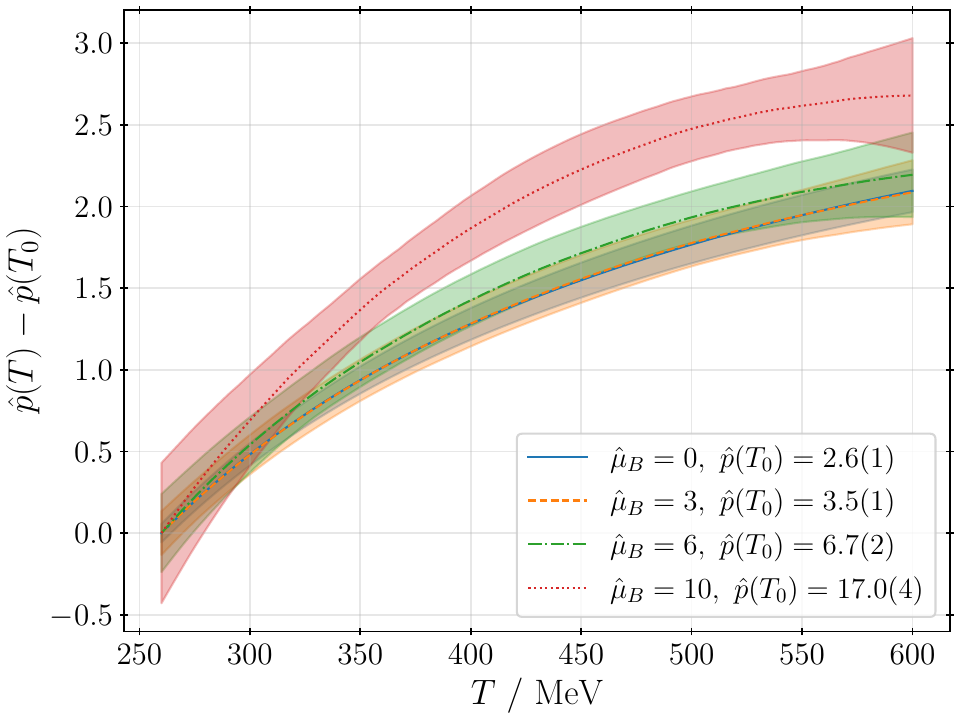}
		\end{subfigure}
	}
	\caption{Similar as \cref{fig:n_B}, but for $\hat{p}$ instead of $\hat{n}_B$.}
	\label{fig:p}
\end{figure}

Since we have direct access to the baryon density as a function of the chemical potential, we may use it compute the pressure difference with respect to $\mu_B=0$,
\begin{equation}\label{eq:delta_p}
	\Delta\hat{p}(T,\mu_B) = \hat{p}(T,\mu_B) - \hat{p}(T,0)\;,
\end{equation}
via simple numerical integration according to \cite{Sex19}
\begin{equation}
	\Delta\hat{p}(T,\mu_B) = \frac{1}{T}\int_0^{\mu_B}d\mu_B'\hat{n}_B(T,\mu_B')\;.
\end{equation}
With this, we can then compute $\hat{p}(T,\mu_B)$ using \eqref{eq:delta_p}, where we take $\hat{p}(T,0)$ from (the Supplementary Information of) \cite{BFG16}. The $\hat{\mu}_B$-dependence of $\hat{p}$ for different temperatures is shown in \cref{fig:p} (left). We observe a strong dependence on $\hat{\mu}_B$ and a milder (albeit non-negligible) dependence on $T$. The latter is further substantiated in \cref{fig:p} (right), where we show the subtracted dimensionless pressure $\hat{p}(T)-\hat{p}(T_0)$ as a function of $T$ for different $\hat{\mu}_B$.

\begin{figure}[t]
	\centerline{
		\begin{subfigure}{.49\linewidth}
			\includegraphics[scale=0.48]{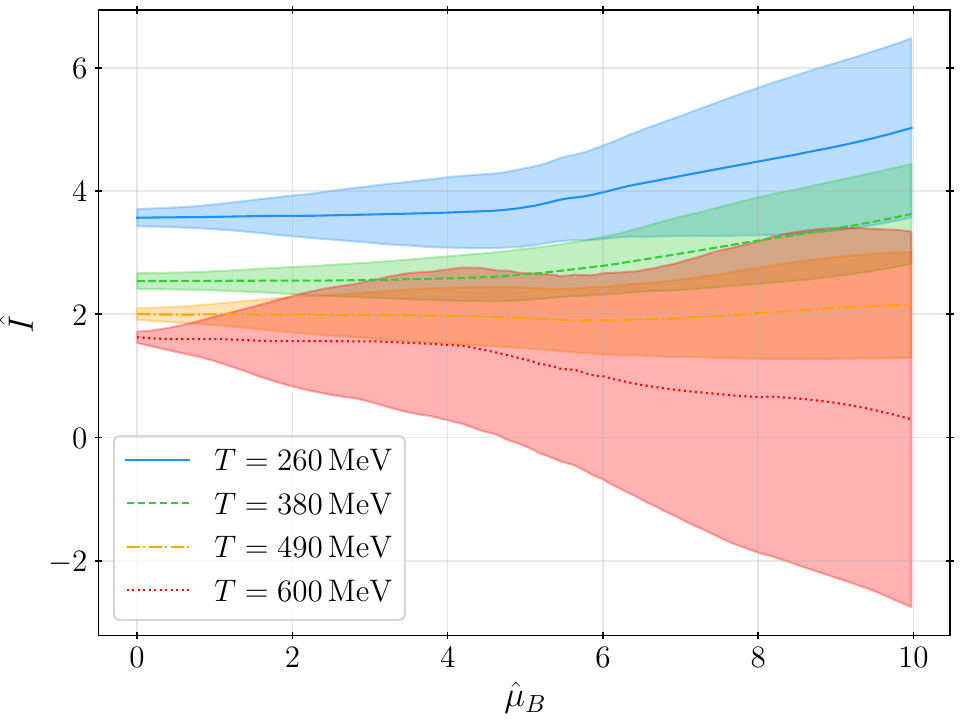}
		\end{subfigure}
		\begin{subfigure}{.49\linewidth}
			\includegraphics[scale=0.48]{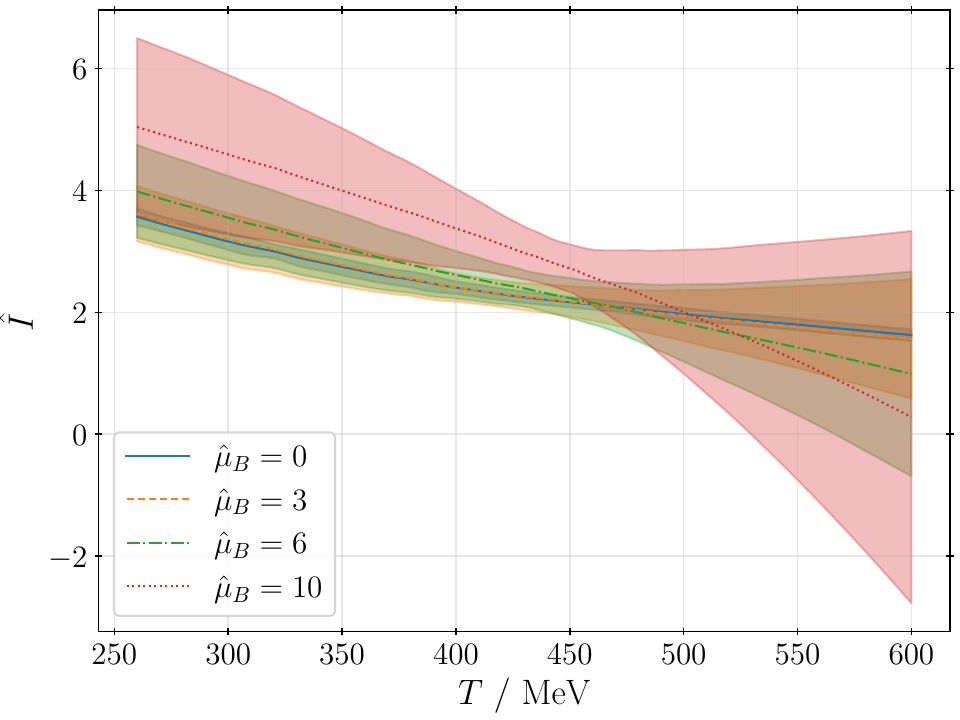}
		\end{subfigure}
	}
	\caption{Dependence of the dimensionless trace anomaly on $\hat{\mu}_B$ for different $T$ \emph{(left)} and its dependence on $T$ for different $\hat{\mu}_B$ \emph{(right)}.}
	\label{fig:I}
\end{figure}

The pressure difference $\Delta\hat{p}(T,\mu_B)$ now allows for the computation of the trace anomaly difference, $\Delta\hat{I}(T,\mu_B) = \hat{I}(T,\mu_B) - \hat{I}(T,0)$, via the $T$-derivative
\begin{equation}
	\hat{I}(T,\mu_B) = T\frac{\partial\Delta\hat{p}(T,\mu_B)}{\partial T}\bigg\vert_{\hat{\mu}_B=const.}\;,
\end{equation}
which we evaluate numerically after fitting our data in $T$. Note that we do not keep the volume constant while evaluating the derivative, but we expect the resulting error to be negligible. To be sure, we have compared the result of this procedure with an alternative approach based on the computation of renormalized expectation values for both the gauge action and the chiral condensates (see, e.g., \cite{Phi13r}) and found consistent results within the statistical uncertainties. In order to obtain $\hat{I}(T,\mu_B)$, we then again add the contribution at $\mu_B=0$, which we take from (the Supplementary Information of) \cite{BFG16} as well. Our results for the dimensionless trace anomaly as a function of $\hat{\mu}_B$ for different temperatures are shown in \cref{fig:I} (left). While the combined statistical and systematic uncertainties are too large to make quantitative conclusions in this case, our data suggest an increase of $\hat{I}$ with $\hat{\mu}_B$ at least for lower temperatures. Finally, the $T$-dependence of $\hat{I}$ for different $\hat{\mu}_B$ is shown in \cref{fig:I} (right), which suggests an overall decrease of $\hat{I}$ with the temperature. With the trace anomaly and the pressure, one may compute additional thermodynamic quantities relevant for the equation of state, such as the energy density, entropy density, enthalpy density, specific heat capacity, or the speed of sound. The computation of these quantities, however, is deferred to a future publication.

We would like to conclude with a brief comment regarding the practical limitations of our approach, in particular the restriction to high temperatures. As was realized in \cite{NS15}, singularities in the complex Langevin drift term, as they appear in lattice QCD due to the logarithm of the fermion determinant, can be problematic if they are probed by the stochastic evolution. Indeed, according to our current understanding, poles of the drift term are the main limitation preventing us from reaching lower temperatures in our setting. It is tempting to interpret this breakdown to be caused by zero modes of the Dirac operator, i.e., to have a topological origin. However, since, on the one hand, we work with non-unitary gauge links and, on the other hand, it is the full staggered fermion matrix (including mass terms and chemical potentials) that enters the determinant, such an interpretation is likely incorrect or at least incomplete. In particular, the role of topology in the non-unitary domain is not fully understood. Furthermore, it should be mentioned that the lowest temperature we can reach within our continuum extrapolation as of now, i.e., $T\approx260\,\mathrm{MeV}$, appears to be roughly independent of the chemical potential.

	\section{Summary \& conclusions}
	We have presented continuum-extrapolated lattice results for the baryon density, pressure, and trace anomaly of QCD at the physical point in the region of high temperature and baryon density using the complex Langevin equation. The ratios of baryon chemical potential to temperature we can reach in our simulations are unprecedented for continuum-extrapolated lattice approaches with a physical pion mass and we work in a strangeness- and charm-neutral setting with isospin symmetry. While our simulations are so far restricted to temperatures well above the QCD crossover, they provide a useful complimentary view on the QCD phase diagram, elucidating previously uncharted regions. In particular, our results might be useful in hydrodynamics applications. 
	
	\section*{Acknowledgments}
		The authors thank Astrid Carstensen, Erhard Seiler, and Ion-Olimpiu Stamatescu for discussions on a regular basis and collaboration on related topics. We are indebted to Szabolcs Borsányi and Jana Günther for providing us with the data for vanishing chemical potential and for the LCP. Moreover, M.M.\ is grateful for enlightening discussions with Szabolcs Borsányi, Matteo Giordano, and Tamás Kovács. This research was funded in whole or in part by the Austrian Science Fund (FWF) \href{https://doi.org/10.55776/P36875}{10.55776/P36875}. Support by the Dr.\ Heinrich Jörg-Stiftung, Karl-Franzens-Universität Graz, Austria, is acknowledged. The computational results have been obtained via the Austrian Scientific Computing (ASC) infrastructure and the computing cluster of the university of Graz (GSC).
	
	\bibliographystyle{JHEP}
	\bibliography{bibliography}

\end{document}